\documentclass[11pt,reprint,english,aps,prl,tightenlines,superscriptaddress,showpacs,showkeys,twocolumn,nobalancelastpage]{revtex4-1}
\usepackage[T1]{fontenc}
\usepackage[latin1]{inputenc}
\usepackage{graphicx}
\usepackage{amsfonts}
\usepackage{amssymb}
\usepackage{subfigure}
\usepackage{setspace}
\usepackage{mathtools}

\usepackage{latexsym}

\usepackage{babel}
\begin{document}

\title{Simple model squirmers with tunable velocity}

\author{Shashi Thutupalli}
\email{shashi.thutupalli@ds.mpg.de}
\affiliation{Max Planck Institute for Dynamics and Self-Organization, Bunsenstr. 10, 37073 G\"ottingen, Germany}
\author{Ralf Seemann}
\affiliation{Max Planck Institute for Dynamics and Self-Organization, Bunsenstr. 10, 37073 G\"ottingen, Germany}
\affiliation{Experimental Physics, Saarland University, Saarbr\"ucken, Germany}
\author{Stephan Herminghaus}
\email{stephan.herminghaus@ds.mpg.de}
\affiliation{Max Planck Institute for Dynamics and Self-Organization, Bunsenstr. 10, 37073 G\"ottingen, Germany}


\begin{abstract}
We present a scheme of self-propelling liquid droplets which closely mimics the locomotion of some protozoal organisms, so-called squirmers. In contrast to other schemes proposed earlier, locomotion paths are not self-avoiding, since the effect of the squirmer on the surrounding medium is weak. Our results suggest that not only the velocity, but also the mode of operation (i.e., the spherical harmonics of the flow field) can be controlled by appropriate variation of parameters.
\end{abstract}
  

\maketitle

Microscopic organisms have developed a variety of strategies to achieve locomotion in liquids at low Reynolds' numbers \cite{Lauga2009}. Flagellated cells such as certain bacteria, sperm, $\it{E.Coli}$, and trypanosomes use nonreciprocal motions of their flagella to overcome viscous forces and set themselves in motion \cite{Berg1973, *Friedrich2010, *Bray2000, *Rodriguez2009}. Organisms such as cyanobacteria, paramecium, and Volvox belong to a class of swimmers referred to as squirmers and are driven by tangential and/or radial deformations of the cell surface \cite{Ehlers1996, *Ishikawa2006, *Drescher2009}. Squirming motion is very appealing for elucidating the hydrodynamics of microscale swimming, since the velocities in the near and far field around such a swimming organism can be well described analytically \cite{Blake1971, Downton2009}. This makes such swimmers ideally suited for the quantitative study of many open questions regarding the hydrodynamic effects on their interaction with surfaces and with each other, their behavior in external flow, and the origins of coupled and collective behavior \cite{Drescher2009, Ishikawa2006}. However, in contrast to the large number of natural micro-swimmers and analytical models, very few artificial self propelled objects have been realized \cite{Howse2007, *Paxton2004, *Toyota2009, *Hanczyc2007}, and even fewer mimic actual biological mechanisms \cite{Dreyfus2005}. Apart from potential technological benefits, a significant appeal of model systems is the ability to provide quantitative control and minimize the individual variations pertinent to biological systems.

We present here a simple model squirmer consisting of an aqueous droplet moving in an oil 'background' phase. Propulsion arises due to the spontaneous bromination of mono-olein (\textit{rac}-Glycerol-1-Mono-oleate) as a surfactant, which is abundantly present in the oil phase, such that the droplet interface is covered by a dense surfactant monolayer. The bromine 'fuel' is supplied from inside the droplet, such that bromination proceeds mainly at the droplet surface. It results in saturation of the unsaturated C-C bond in the alkyl chain of the surfactant, thus rendering it a weaker surfactant. As we will see below, this results in a self-sustained bromination gradient along the drop surface, which propels the droplet due to Marangoni stresses. 

\begin{figure}[h]
        \includegraphics[width=\columnwidth]{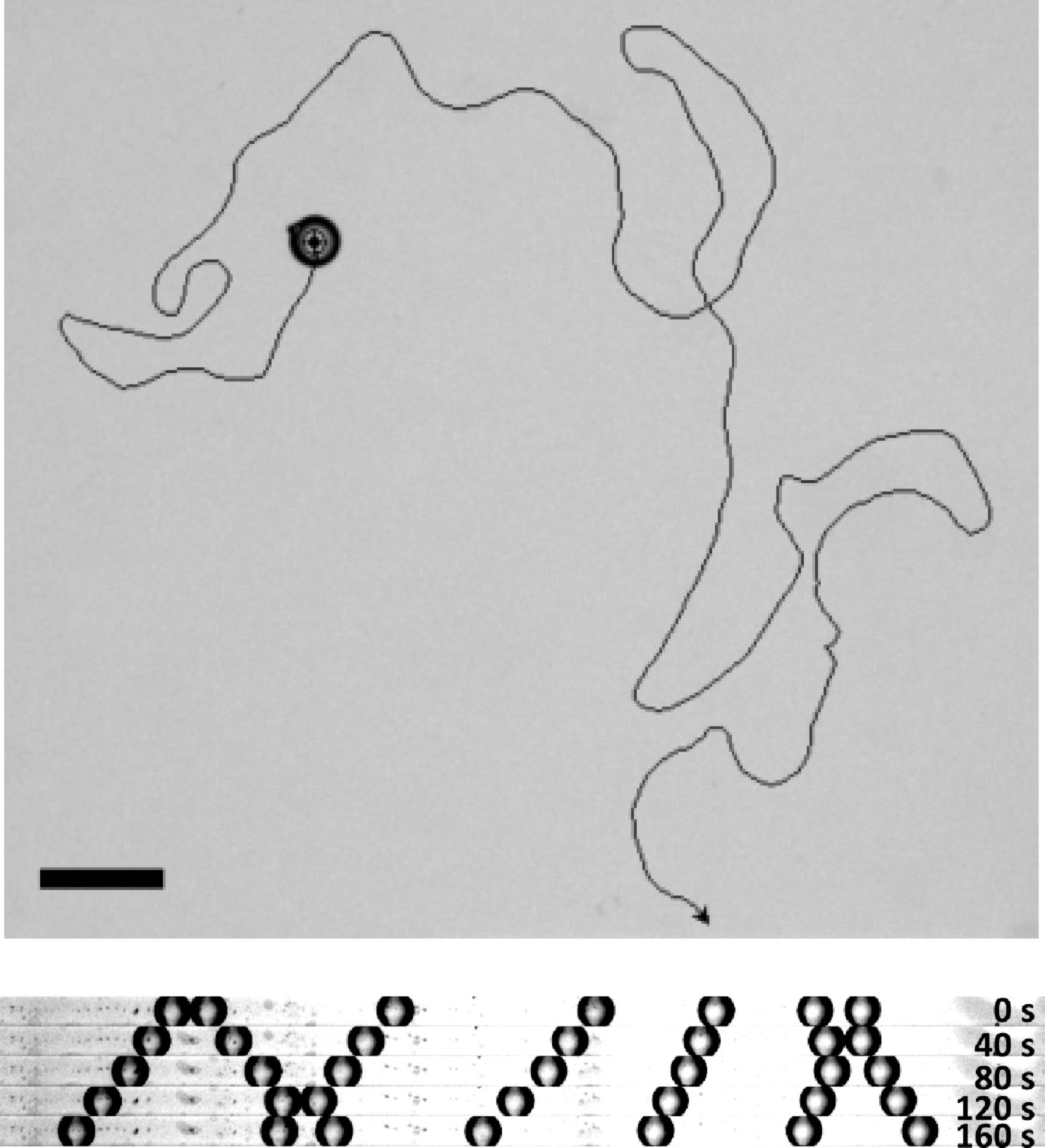}
    \caption{Top: Path of a single squirmer droplet. The persistence length is clearly large compared to the droplet radius, indicating propelled motion (Scale bar: 300 microns). Bottom: Time lapse series of seven droplets in a mocrochannel. The droplets change direction without noticeable reduction in velocity.}
    \label{Squirmerpath}
\end{figure}

We demonstrate this squirmer scheme with nanoliter droplets containing 25 mM bromine water in a continuous oil phase of squalane containing 50 mM mono-olein (MO). The critical  micelle concentration (CMC) is 1.5 mM. The droplets are confined by two hydrophobic glass plates to a quasi 2 dimensional space, thus simplifying droplet tracking. The top panel of Fig.~\ref{Squirmerpath} shows the trajectory of a single squirmer droplet over a duration of 400 seconds. The velocity of the droplet in this duration is roughly constant at about 15 $\mu$m/sec. The trajectory is reminiscent of a random walk, with a persistence length which is larger than the droplet size and clearly far beyond what would be expected for Brownian motion. A particularly important observation is that the trajectory crosses itself. This is in sharp contrast to other schemes, where the propulsion mechanism itself changes the surrounding medium strongly enough to prevent self-crossing of the path \cite{Ondarcuhu1995}. That this is clearly not the case here is demonstrated even more convincingly in the bottom panel, which shows a time lapse representation of seven droplets moving in a micro-channel. As two drops touch each other, they reverse their direction of motion and perambulate the channel again, without significant reduction in velocity.

\begin{figure}[h]
        \includegraphics[width=\columnwidth]{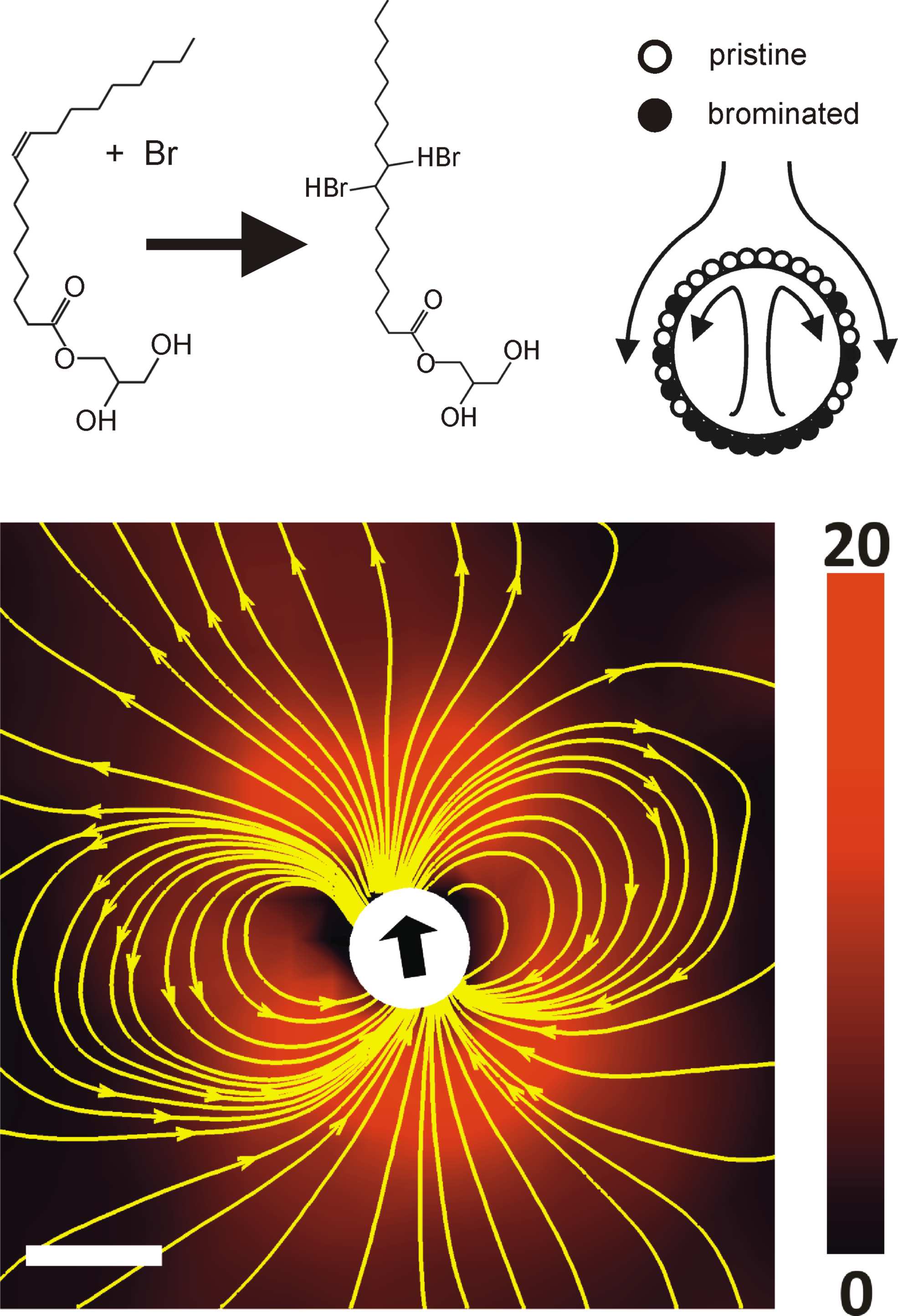}
    \caption{Top: Schematic of a micro-droplet squirmer. Bromination increases the tension of the droplet surface from 1.8 mN/m to 3.1 mN/m. The convective flow pattern (shown in the rest frame of the droplet) is accompanied by a gradient in the bromination density. The corresponding Marangoni stress propels the droplet. Bottom: velocity field around a droplet squirmer. The magnitude of the flow velocity (color code) and streamlines along a horizontal section through the center of a squirmer droplet are shown. Scale bar: 100 microns; velocity scale (right) in microns per second.}
    \label{Flowprofilescheme}
\end{figure}

In order to gain some insight into the propulsion mechanism, let us consider a spherical droplet with radius $R$. The total coverage, $c$, of the droplet surface with the mono-olein, either brominated or not, is assumed to be roughly constant and in equilibrium with the micellar phase in the oil. The brominated fractional coverage shall be called $b$. If the droplet moves, there is (in the rest frame of the droplet) an axisymmetric flow field, $u(\theta)$, along its surface. The equation of motion for $b$ is 
\begin{equation}
\frac{\partial b}{\partial t} = 
 k (b_0 - b) 
+ {\rm div}
\left(
D_i {\rm grad}b - u b \right)
\label{EqMot:bromination}
\end{equation}
where $b_0$ is the equilibrium coverage with brominated mono-olein (brMO). It is determined by the bromine supply from inside the droplet and the rate constant, $k$, of escape of brMO into the oil phase. $D_i$ is the diffusivity of the surfactant within the interface. 

The droplet motion is accompanied by a flow pattern within the droplet and in the neighboring oil, which can be determined from $u(\theta)$ \cite{Levan1976,Levan1981}. The corresponding viscous tangential stress exerted on the drop surface must be balanced by the Marangoni stress, ${\rm grad} \gamma(\theta) = M {\rm grad}b(\theta)$, where $\gamma$ is the surface tension of the surfactant-laden oil/water interface, and $M = d \gamma/d b$ is the Marangoni coefficient of the system. Expanding the bromination density in spherical harmonics, 
\begin{equation}
b(\theta) =  \sum_{m=0}^{\infty } b_m P_m (\cos \theta)
\end{equation}
we can express the velocity field \cite{Levan1976, *Levan1981} at the interface as
\begin{equation}
u (\theta)= \frac{M}{\mu \sin\theta}\sum_{m=1}^{\infty}\frac{m(m+1)b_{m}C_{m+1}^{-1/2}(\cos\theta)}{2m+1}
\end{equation}
where $C_n^{\alpha}$ denote Gegenbauer polynomials, and $\mu$ is the sum of the liquid viscosities outside and inside the droplet. Inserting this into eq.~(\ref{EqMot:bromination}) and exploiting the orthogonality relations of Gegenbauer and Legendre polynomials, we obtain
\begin{equation}
\frac{d b_m}{dt} = \left[ m(m+1) \left(\frac{b_0 M}{(2m+1)R\mu} - \frac{D_i}{R^2}\right) - k \right] b_m
\label{Eq:GrowthRate}
\end{equation}
for all $m > 0$. We see that the different modes decouple, as far as linear stability is concerned. As long as $b_0 M$ is small enough, the resting state is stable against fluctuations. However, when $b_0 M$ exceeds a critical value, the resting state is unstable, and the droplet spontaneously starts to move. It is straightforward to see that for $k < 3D_i
/R^2$, this happens first for the lowest mode at $m=1$ which corresponds to a purely dipolar flow field. 

In order to determine the flow profile around a squirming drop, we performed particle image velocimetry using a standard setup (ILA GmbH). The oil phase is seeded with 200 nm green fluorescent polystyrene beads (Duke Scientific) to be used as tracers. The result in the bottom panel of Fig.~\ref{Flowprofilescheme}, shows that the flow profile indeed resembles a purely dipolar field. We note, however, that by careful adjustment of conditions, such as adding suitable solvents to the oil phase, one should be able to tune $k$, and thus control which modes become unstable, thereby adjusting the flow pattern in the droplet and its vicinity \cite{Downton2009}. 
 
To discuss the steady state velocity, $V$, we reconsider the total surfactant coverage, $c$. This adjusts itself as a balance between the molecular adsorption energy at the water/oil interface and the mutual repulsion of the adsorbed surfactant molecules. The equilibrium coverage thus represents a minimum in the interfacial energy. As a consequence, the interfacial tension will not change to first order if $c$ is varied. Deviations of $c$ from its equilibrium value, which come about necessarily for any finite $V$, will thus be replenished from the surrounding oil phase without noticeable Marangoni stresses. 
  
For a dipolar flow pattern as shown in Fig.~\ref{Flowprofilescheme}, we have ${\rm div} u(\theta) \propto \cos \theta$. The density of surfactant thus takes the form $\rho \approx \rho_0 + \delta\rho f(r) \cos\theta$, where $\delta\rho \propto V$. $f(r)$ is some function of the radius. If $D_m$ is the diffusivity of the micelles, the diffusion current,  $D_m \partial\rho/\partial r$, must balance the depletion rate at the drop interface, $c {\rm div} u(\theta) = \frac{3Vc}{2R}\cos\theta$ \cite{Downton2009}. As long as this holds, the only source of appreciable Marangoni stresses is the gradient in bromination density, $b(\theta)$. The drop thus keeps taking up speed according to eq.~(\ref{Eq:GrowthRate}). This comes to an end when $\delta\rho \approx \rho_0$. The surfactant layer at the leading end of the drop surface can then not be replenished anymore, and $c$ comes substantially below its equilibrium value. This leads to an increase of surface tension accompanied by a  'backward' Marangoni stress, and thus finally to a saturation of the velocity. According to the reasoning above, we expect that $V \approx 2 \rho_0 D_m/3c$.
 
There is no literature value for the diffusivity of MO micelles in squalane, but we can estimate it on the basis of the Stokes-Einstein relation assuming the radius of the micelles to be similar to the length of a MO molecule (2.3 nm). Using 36 mPa.s for the viscosity of the squalane, we obtain $D_m =$ 2.6x$10^{-12} {\rm m}^2$/s. We thus predict $V/\rho_0 \approx 0.27 \  \frac{\mu{\rm m/sec}}{{\rm mM/l}}$. 
 
\begin{figure}[h!]
        \includegraphics[width=\columnwidth]{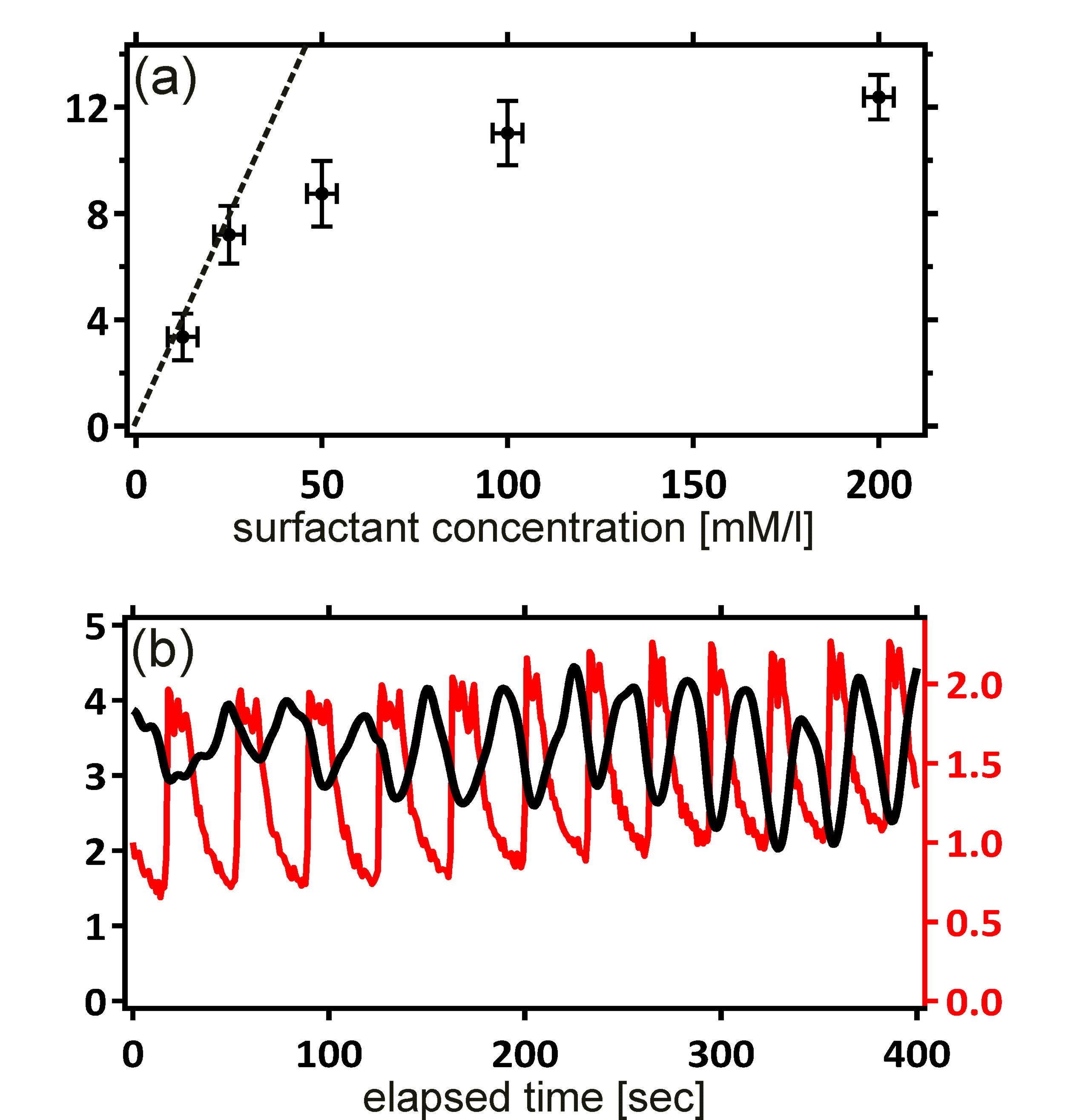}
    \caption{Squirmer velocity in units of microns per second, as a function of (a) surfactant concentration, where each point is an average of 50 different squirmer droplets each of diameter $\sim 80$ microns and (b) time when an oscillating chemical reaction (BZ) takes place in the droplet. For the latter case, the optical transmission of the droplet is plotted in red (arbitrary units, linear scale) along with the velocity trace.} 
    \label{VelocitySurfactantOscillations}
\end{figure}

This can be tested experimentally. In order to measure $V$, we used a reaction mixture within the droplets similar to the Belousov-Zhabotinski (BZ) reaction, with reactant concentrations adjusted such as to prevent chemical oscillations. This results in a spatially and temporally constant bromine release rate in the aqueous phase for an extended period of time. Fig.~\ref{VelocitySurfactantOscillations}a shows the dependence of the droplet velocity as a function of the mono-olein concentration in the oil phase.  The initial linear increase is in agreement with the above prediction (dashed line). As the surfactant density is further increased (and thus the velocity), the complex exchange processes between the water/oil interface and the micelles will finally become the rate limiting step. As a consequence, the velocity is expected to level off, in agreement with our data.

Our model predicts the locomotion velocity to be independent of the bromine release rate. This can of course not be strictly true in general, and it is instructive to demonstrate this experimentally. We take advantage again of the BZ reaction, this time creating an oscillating bromine concentration inside the droplet. This oscillation can be easily visualized by the optical transmission of the droplet. Care was taken that there were no spatio-temporal patterns within the droplets, in contrast to some other recent work on droplet locomotion \cite{Kitahata2011}. Fig.~\ref{VelocitySurfactantOscillations}b shows an overlay of the the BZ chemical oscillations and the velocity of the squirmer droplet, which are anti-phase with each other. As it is known for the BZ reaction, the decrease of the transmitted intensity corresponds to an increase in the bromine concentration within the droplet. It is obviously possible to control the droplet velocity in some range, and we have demonstrated here how this can be done even in an autonomous manner. We expect this model squirmer scheme to be instrumental in elucidating the hydrodynamics of collective behaviour, in particular concerning locomoting organisms.

Inspiring discussions with Jürgen Vollmer, Martin Brinkmann, and Daniel Herde are gratefully acknowledged. This work has been supported within the Collaborative Research Programme SFB 755 (funded by the German Science Foundation).

\begin{thebibliography}{19}%
\makeatletter
\providecommand \@ifxundefined [1]{%
 \@ifx{#1\undefined}
}%
\providecommand \@ifnum [1]{%
 \ifnum #1\expandafter \@firstoftwo
 \else \expandafter \@secondoftwo
 \fi
}%
\providecommand \@ifx [1]{%
 \ifx #1\expandafter \@firstoftwo
 \else \expandafter \@secondoftwo
 \fi
}%
\providecommand \natexlab [1]{#1}%
\providecommand \enquote  [1]{``#1''}%
\providecommand \bibnamefont  [1]{#1}%
\providecommand \bibfnamefont [1]{#1}%
\providecommand \citenamefont [1]{#1}%
\providecommand \href@noop [0]{\@secondoftwo}%
\providecommand \href [0]{\begingroup \@sanitize@url \@href}%
\providecommand \@href[1]{\@@startlink{#1}\@@href}%
\providecommand \@@href[1]{\endgroup#1\@@endlink}%
\providecommand \@sanitize@url [0]{\catcode `\\12\catcode `\$12\catcode
  `\&12\catcode `\#12\catcode `\^12\catcode `\_12\catcode `\%12\relax}%
\providecommand \@@startlink[1]{}%
\providecommand \@@endlink[0]{}%
\providecommand \url  [0]{\begingroup\@sanitize@url \@url }%
\providecommand \@url [1]{\endgroup\@href {#1}{\urlprefix }}%
\providecommand \urlprefix  [0]{URL }%
\providecommand \Eprint [0]{\href }%
\@ifxundefined \urlstyle {%
  \providecommand \doi  [0]{\begingroup \@sanitize@url \@doi}%
  \providecommand \@doi [1]{\endgroup \@@startlink {\doibase
  #1}doi:\discretionary {}{}{}#1\@@endlink }%
}{%
  \providecommand \doi  [0]{doi:\discretionary{}{}{}\begingroup
  \urlstyle{rm}\Url }%
}%
\providecommand \doibase [0]{http://dx.doi.org/}%
\providecommand \Doi [0]{\begingroup \@sanitize@url \@Doi }%
\providecommand \@Doi  [1]{\endgroup\@@startlink{\doibase#1}\@@Doi}%
\providecommand \@@Doi [1]{#1\@@endlink}%
\providecommand \selectlanguage [0]{\@gobble}%
\providecommand \bibinfo  [0]{\@secondoftwo}%
\providecommand \bibfield  [0]{\@secondoftwo}%
\providecommand \translation [1]{[#1]}%
\providecommand \BibitemOpen [0]{}%
\providecommand \bibitemStop [0]{}%
\providecommand \bibitemNoStop [0]{.\EOS\space}%
\providecommand \EOS [0]{\spacefactor3000\relax}%
\providecommand \BibitemShut  [1]{\csname bibitem#1\endcsname}%
\bibitem [{\citenamefont {Lauga}\ and\ \citenamefont
  {Powers}(2009)}]{Lauga2009}%
  \BibitemOpen
  \bibfield  {author} {\bibinfo {author} {\bibfnamefont {E.}~\bibnamefont
  {Lauga}}\ and\ \bibinfo {author} {\bibfnamefont {T.~R.}\ \bibnamefont
  {Powers}},\ }\href@noop {} {\bibfield  {journal} {\bibinfo  {journal} {Rep.
  Prog. Phys.},\ }\textbf {\bibinfo {volume} {72}},\ \bibinfo {pages} {096601}
  (\bibinfo {year} {2009})}\BibitemShut {NoStop}%
\bibitem [{\citenamefont {Berg}\ and\ \citenamefont
  {Anderson}(1973)}]{Berg1973}%
  \BibitemOpen
  \bibfield  {author} {\bibinfo {author} {\bibfnamefont {H.~C.}\ \bibnamefont
  {Berg}}\ and\ \bibinfo {author} {\bibfnamefont {R.~A.}\ \bibnamefont
  {Anderson}},\ }\href@noop {} {\bibfield  {journal} {\bibinfo  {journal}
  {Nature},\ }\textbf {\bibinfo {volume} {245}},\ \bibinfo {pages} {380}
  (\bibinfo {year} {1973})}\BibitemShut {NoStop}%
\bibitem [{\citenamefont {Friedrich}\ \emph {et~al.}(2010)\citenamefont
  {Friedrich}, \citenamefont {Riedel-Kruse}, \citenamefont {Howard},\ and\
  \citenamefont {J\"{u}licher}}]{Friedrich2010}%
  \BibitemOpen
  \bibfield  {author} {\bibinfo {author} {\bibfnamefont {B.~M.}\ \bibnamefont
  {Friedrich}}, \bibinfo {author} {\bibfnamefont {I.~H.}\ \bibnamefont
  {Riedel-Kruse}}, \bibinfo {author} {\bibfnamefont {J.}~\bibnamefont
  {Howard}}, \ and\ \bibinfo {author} {\bibfnamefont {F.}~\bibnamefont
  {J\"{u}licher}},\ }\href@noop {} {\bibfield  {journal} {\bibinfo  {journal}
  {J. Exp. Biol.},\ }\textbf {\bibinfo {volume} {213}},\ \bibinfo {pages}
  {1226} (\bibinfo {year} {2010})}\BibitemShut {NoStop}%
\bibitem [{\citenamefont {Bray}(2000)}]{Bray2000}%
  \BibitemOpen
  \bibfield  {author} {\bibinfo {author} {\bibfnamefont {D.}~\bibnamefont
  {Bray}},\ }\href@noop {} {\emph {\bibinfo {title} {Cell Movements}}}\
  (\bibinfo  {publisher} {New York: Garland},\ \bibinfo {year}
  {2000})\BibitemShut {NoStop}%
\bibitem [{\citenamefont {Rodr\'{\i}guez}\ \emph {et~al.}(2009)\citenamefont
  {Rodr\'{\i}guez} \emph {et~al.}}]{Rodriguez2009}%
  \BibitemOpen
  \bibfield  {author} {\bibinfo {author} {\bibfnamefont {J.~A.}\ \bibnamefont
  {Rodr\'{\i}guez}} \emph {et~al.},\ }\href@noop {} {\bibfield  {journal}
  {\bibinfo  {journal} {Proc. Natl. Acad. Sci. U. S. A.},\ }\textbf {\bibinfo
  {volume} {106}},\ \bibinfo {pages} {19322} (\bibinfo {year}
  {2009})}\BibitemShut {NoStop}%
\bibitem [{\citenamefont {Ehlers}\ \emph {et~al.}(1996)\citenamefont {Ehlers},
  \citenamefont {Samuel}, \citenamefont {Berg},\ and\ \citenamefont
  {Montgomery}}]{Ehlers1996}%
  \BibitemOpen
  \bibfield  {author} {\bibinfo {author} {\bibfnamefont {K.~M.}\ \bibnamefont
  {Ehlers}}, \bibinfo {author} {\bibfnamefont {D.}~\bibnamefont {Samuel}},
  \bibinfo {author} {\bibfnamefont {H.~C.}\ \bibnamefont {Berg}}, \ and\
  \bibinfo {author} {\bibfnamefont {R.}~\bibnamefont {Montgomery}},\
  }\href@noop {} {\bibfield  {journal} {\bibinfo  {journal} {Proc. Natl. Acad.
  Sci. U. S. A.},\ }\textbf {\bibinfo {volume} {93}},\ \bibinfo {pages} {8340}
  (\bibinfo {year} {1996})}\BibitemShut {NoStop}%
\bibitem [{\citenamefont {Ishikawa}\ and\ \citenamefont
  {Hota}(2006)}]{Ishikawa2006}%
  \BibitemOpen
  \bibfield  {author} {\bibinfo {author} {\bibfnamefont {T.}~\bibnamefont
  {Ishikawa}}\ and\ \bibinfo {author} {\bibfnamefont {M.}~\bibnamefont
  {Hota}},\ }\href@noop {} {\bibfield  {journal} {\bibinfo  {journal} {J. Exp.
  Biol.},\ }\textbf {\bibinfo {volume} {209}},\ \bibinfo {pages} {4452}
  (\bibinfo {year} {2006})}\BibitemShut {NoStop}%
\bibitem [{\citenamefont {Drescher}\ \emph {et~al.}(2009)\citenamefont
  {Drescher} \emph {et~al.}}]{Drescher2009}%
  \BibitemOpen
  \bibfield  {author} {\bibinfo {author} {\bibfnamefont {K.}~\bibnamefont
  {Drescher}} \emph {et~al.},\ }\href@noop {} {\bibfield  {journal} {\bibinfo
  {journal} {Phys. Rev. Lett.},\ }\textbf {\bibinfo {volume} {102}},\ \bibinfo
  {pages} {1} (\bibinfo {year} {2009})}\BibitemShut {NoStop}%
\bibitem [{\citenamefont {Blake}(1971)}]{Blake1971}%
  \BibitemOpen
  \bibfield  {author} {\bibinfo {author} {\bibfnamefont {J.~R.}\ \bibnamefont
  {Blake}},\ }\href@noop {} {\bibfield  {journal} {\bibinfo  {journal} {J.
  Fluid Mech.},\ }\textbf {\bibinfo {volume} {46}},\ \bibinfo {pages} {199}
  (\bibinfo {year} {1971})}\BibitemShut {NoStop}%
\bibitem [{\citenamefont {Downton}\ and\ \citenamefont
  {Stark}(2009)}]{Downton2009}%
  \BibitemOpen
  \bibfield  {author} {\bibinfo {author} {\bibfnamefont {M.~T.}\ \bibnamefont
  {Downton}}\ and\ \bibinfo {author} {\bibfnamefont {H.}~\bibnamefont
  {Stark}},\ }\href@noop {} {\bibfield  {journal} {\bibinfo  {journal} {J.
  Phys.: Condens. Matter},\ }\textbf {\bibinfo {volume} {21}},\ \bibinfo
  {pages} {204101} (\bibinfo {year} {2009})}\BibitemShut {NoStop}%
\bibitem [{\citenamefont {Howse}\ \emph {et~al.}(2007)\citenamefont {Howse}
  \emph {et~al.}}]{Howse2007}%
  \BibitemOpen
  \bibfield  {author} {\bibinfo {author} {\bibfnamefont {J.}~\bibnamefont
  {Howse}} \emph {et~al.},\ }\href@noop {} {\bibfield  {journal} {\bibinfo
  {journal} {Phys. Rev. Lett.},\ }\textbf {\bibinfo {volume} {99}} (\bibinfo
  {year} {2007})}\BibitemShut {NoStop}%
\bibitem [{\citenamefont {Paxton}\ \emph {et~al.}(2004)\citenamefont {Paxton}
  \emph {et~al.}}]{Paxton2004}%
  \BibitemOpen
  \bibfield  {author} {\bibinfo {author} {\bibfnamefont {W.~F.}\ \bibnamefont
  {Paxton}} \emph {et~al.},\ }\href@noop {} {\bibfield  {journal} {\bibinfo
  {journal} {J. Am. Chem. Soc.},\ }\textbf {\bibinfo {volume} {126}},\ \bibinfo
  {pages} {13424} (\bibinfo {year} {2004})}\BibitemShut {NoStop}%
\bibitem [{\citenamefont {Toyota}\ \emph {et~al.}(2009)\citenamefont {Toyota}
  \emph {et~al.}}]{Toyota2009}%
  \BibitemOpen
  \bibfield  {author} {\bibinfo {author} {\bibfnamefont {T.}~\bibnamefont
  {Toyota}} \emph {et~al.},\ }\href@noop {} {\bibfield  {journal} {\bibinfo
  {journal} {J. Am. Chem. Soc.},\ }\textbf {\bibinfo {volume} {131}},\ \bibinfo
  {pages} {5012} (\bibinfo {year} {2009})}\BibitemShut {NoStop}%
\bibitem [{\citenamefont {Hanczyc}\ \emph {et~al.}(2007)\citenamefont {Hanczyc}
  \emph {et~al.}}]{Hanczyc2007}%
  \BibitemOpen
  \bibfield  {author} {\bibinfo {author} {\bibfnamefont {M.~M.}\ \bibnamefont
  {Hanczyc}} \emph {et~al.},\ }\href@noop {} {\bibfield  {journal} {\bibinfo
  {journal} {J. Am. Chem. Soc.},\ }\textbf {\bibinfo {volume} {129}},\ \bibinfo
  {pages} {9386} (\bibinfo {year} {2007})}\BibitemShut {NoStop}%
\bibitem [{\citenamefont {Dreyfus}\ \emph {et~al.}(2005)\citenamefont {Dreyfus}
  \emph {et~al.}}]{Dreyfus2005}%
  \BibitemOpen
  \bibfield  {author} {\bibinfo {author} {\bibfnamefont {R.}~\bibnamefont
  {Dreyfus}} \emph {et~al.},\ }\href@noop {} {\bibfield  {journal} {\bibinfo
  {journal} {Nature},\ }\textbf {\bibinfo {volume} {437}},\ \bibinfo {pages}
  {862} (\bibinfo {year} {2005})}\BibitemShut {NoStop}%
\bibitem [{\citenamefont {Dos~Santos}\ and\ \citenamefont
  {Ondarcuhu}(1995)}]{Ondarcuhu1995}%
  \BibitemOpen
  \bibfield  {author} {\bibinfo {author} {\bibfnamefont {F.~D.}\ \bibnamefont
  {Dos~Santos}}\ and\ \bibinfo {author} {\bibfnamefont {T.}~\bibnamefont
  {Ondarcuhu}},\ }\href@noop {} {\bibfield  {journal} {\bibinfo  {journal}
  {Phys. Rev. Lett.},\ }\textbf {\bibinfo {volume} {75}},\ \bibinfo {pages}
  {2972} (\bibinfo {year} {1995})}\BibitemShut {NoStop}%
\bibitem [{\citenamefont {Levan}\ and\ \citenamefont
  {Newman}(1976)}]{Levan1976}%
  \BibitemOpen
  \bibfield  {author} {\bibinfo {author} {\bibfnamefont {M.~D.}\ \bibnamefont
  {Levan}}\ and\ \bibinfo {author} {\bibfnamefont {J.}~\bibnamefont {Newman}},\
  }\href@noop {} {\bibfield  {journal} {\bibinfo  {journal} {AIChE J.},\
  }\textbf {\bibinfo {volume} {22}},\ \bibinfo {pages} {695} (\bibinfo {year}
  {1976})}\BibitemShut {NoStop}%
\bibitem [{\citenamefont {Levan}(1981)}]{Levan1981}%
  \BibitemOpen
  \bibfield  {author} {\bibinfo {author} {\bibfnamefont {M.}~\bibnamefont
  {Levan}},\ }\href@noop {} {\bibfield  {journal} {\bibinfo  {journal} {J.
  Colloid Interface Sci.},\ }\textbf {\bibinfo {volume} {83}},\ \bibinfo
  {pages} {11} (\bibinfo {year} {1981})}\BibitemShut {NoStop}%
\bibitem [{\citenamefont {Kitahata}\ \emph {et~al.}()\citenamefont {Kitahata},
  \citenamefont {Yoshinaga}, \citenamefont {Nagai},\ and\ \citenamefont
  {Sumino}}]{Kitahata2011}%
  \BibitemOpen
  \bibfield  {author} {\bibinfo {author} {\bibfnamefont {H.}~\bibnamefont
  {Kitahata}}, \bibinfo {author} {\bibfnamefont {N.}~\bibnamefont {Yoshinaga}},
  \bibinfo {author} {\bibfnamefont {K.~H.}\ \bibnamefont {Nagai}}, \ and\
  \bibinfo {author} {\bibfnamefont {Y.}~\bibnamefont {Sumino}},\ }\href@noop {}
  {\bibinfo  {journal} {arXiv:1012.2755v1}}\BibitemShut {NoStop}%
\end{thebibliography}%









\end{document}